\documentclass[prl,twocolumn]{revtex4}
\usepackage{amsfonts}
\usepackage{amsmath}
\usepackage{amssymb}
\begin{document}

\noindent\textbf{Comment on "Monomer Dynamics in Double- and Single-Stranded DNA Polymers"} \\

Despite the success of the bead-spring models \cite{1} in the description of the dynamics of
polymers some problems remain unsolved. Recently, a new "puzzle" has been reported \cite{2}. In
\cite{2} the motion of individual monomers within polymer coils, single- and double-stranded (ss
and ds) DNA, has been observed using fluorescence correlation spectroscopy. The mean square
displacement (MSD) of the end monomer was interpreted within the Rouse (R) and Zimm (Z) models for
flexible polymers. According to \cite{2}, an agreement with the R model has been found for dsDNA,
while ssDNA followed the expected Z-type kinetics. It is discussed in this Comment that the
kinetics observed for long dsDNA is far from the R one and the macromolecule behaves rather as a Z
polymer. The new puzzle is only due to an improper use of the RZ theory.\\\indent\ The quantity
indicating whether the dynamics is of the Z or R type is the draining parameter. For the polymer
internal modes it can be defined as $h(p)=\tau_{pR}/\tau_{pZ}\propto p^{-1/2}$, where $p$ is the
mode number, for the diffusion of the coil it is given by $D_Z/D_R$. Here $\tau_{pi}$ are the
relaxation times in the corresponding model and $D_i$ the diffusion coefficients of the coil
\cite{1}. The "pure" Z and R models correspond to infinitely large and small $h(1)$, respectively.
For the DNA parameters used in \cite{2}, $h(1)$ is neither small nor large. So, for dsDNA with
23100 base pairs (bp) $h(1)\approx 2$ and $D_Z/D_R\approx 4$, so that the R dynamics cannot be
expected. The experiments thus should be analyzed with no preliminary assumption on a specific
dynamics. In \cite{2} the data were compared to the MSD given by the $t^{1/2}$ and $t^{2/3}$ laws
for the R and Z polymers \cite{1,3}. However, these laws assume continuous distribution of the
internal modes and are only rough approximations (with incorrect time dependence for both long and
short times) to the more general expression, which for the end bead reads
\begin{equation} \label{eq1}
\langle r^2(t)\rangle=6Dt+4N(a/\pi)^2\sum_{p=1}^{\infty}p^{-2}\left[1-\exp(-t/\tau_p\right],
\end{equation}
where $D=D_R+D_Z$ is the Kirkwood diffusion coefficient \cite{1} and the relaxation rates are
$\tau_{p}^{-1}=\tau_{pR}^{-1}+\tau_{pZ}^{-1}$ \cite{3, 4} ($N$ is the number of beads and $a$ the
mean square distance between the neighboring beads along the chain). Equation (1) is valid for the
times $t>>R^2\varrho/\eta$ [4], which is satisfied in [2] ($\rho$ is the solvent density, $\eta$
its viscosity, and $R$ the hydrodynamic radius of the coil). Within the interval from 0.02 to 10
ms (where according to \cite{2} the R behavior was observed for dsDNA) the approximation $\langle
r^2(t)\rangle\approx 6D_R t+\left(8k_BTa^{2}t/\pi^{2}\eta b\right)^{1/2}$ ($b$ is the bead radius)
is inapplicable: for the parameters $N=68$, $a=100$ nm, $b=12.8$ nm \cite{2} the sum in Eq. (1)
constitutes from 8 to 80$\%$ of its $t^{1/2}$ approximation. The "R modes" (with $h(p)<1$, which
for the experiment \cite{2} means $p>4$), contribute only 13$\%$ to $\langle r^2(\infty)\rangle$
due to the internal modes. In the interval 0.02 - 10 ms these modes produce from 95 to 34$\%$ of
the total internal part of MSD. Moreover, the diffusion contribution cannot be neglected as in
Ref. [2]. For example, for 23100 bp DNA in the interval 10 - 200 ms (where the Z behavior of dsDNA
has been identified \cite{2}) the diffusion term contributes from 20 to 60$\%$ to the total
MSD.\\\indent\ Numerical calculations using Eq. (1) show that for the parameters from \cite{2} the
correspondence between the data [2] and the theory is unsatisfactory: neither the R nor the Z
model corresponds to the more general RZ model. The polymer parameters thus should be different.
We have obtained them by fitting the RZ model (1) to the data \cite{2}. The diffusion term was
kept in the consideration. The fit was performed with the conditions $Na<L$ and $b<a/2$. The
contour length $L$ was calculated using the known distance between the base pairs along the chain
(0.34 nm). Fixing $N=78$, which corresponds to the same distance (100 nm) between the beads as in
\cite{2}, we obtained $a=99.1$ nm, $b=49.5$ nm and $h(1)\approx8.6$ that indicates the
(approximate) applicability of the Z model. When the R limit of Eq. (1) was fitted to the data,
the obtained parameters yielded the Kuhn length (defined as $l=Na^2/L$), which was only 30 nm
while the accepted value is about 100 nm, in agreement with the RZ case ($l\approx 98$ nm). This
is one more argument against the validity of the R model for the dsDNA dynamics. In summary, as
distinct from the conclusion in \cite{2}, the RZ (being predominantly Zimm) model should be
preferred. An analogous consideration of the ssDNA data [2] showed that also this polymer follows
essentially the Zimm dynamics (with $h(1)\approx 20$ for the 6700 bases ssDNA), in accordance with
\cite{2}.\\\indent\ We are greatly indebted to O. Krichevsky for providing us with the
experimental data \cite{2}.
This work was supported by VEGA, Slovak Republic.\\\\
\noindent J. Tothova, B. Brutovsky, and V. Lisy\\
Institute of Physics, Jesenna 5, 041 54 Kosice, Slovakia\\[0.3cm]


\begin{thebibliography}{}
\bibitem{1} M. Doi and S.F. Edwards, \textit{The Theory of Polymer Dynamics }(Clarendon, Oxford, 1986).
\bibitem{2} R. Shusterman, S. Alon, T. Gavrinyov, and O. Krichevsky,
Phys. Rev. Lett. \textbf{92}, 048303 (2004).
\bibitem{3} E. Dubois-Violette and P.-G. de Gennes,
Physics \textbf{3}, 181 (1967).
\bibitem{4} V. Lisy, J. Tothova, and A.V. Zatovsky, J. Chem. Phys. \textbf{121}, 10699 (2004).

\end{thebibliography}
\end{document}